\documentclass[11pt]{article}
\usepackage{amsmath,amssymb,epsf,color}
\usepackage{mathrsfs}
\usepackage{amsfonts}

\makeatletter \@addtoreset{equation}{section} \makeatother

\def\ft#1#2{{\textstyle{\frac{\scriptstyle #1}{\scriptstyle #2} } }}
\def\ft#1#2{{\textstyle{\frac{\scriptstyle #1}{\scriptstyle #2} } }}
\def\fft#1#2{{\frac{#1}{#2}}}

\def\0{{\sst{(0)}}}
\def\1{{\sst{(1)}}}
\def\2{{\sst{(2)}}}
\def\3{{\sst{(3)}}}
\def\4{{\sst{(4)}}}
\def\5{{\sst{(5)}}}
\def\6{{\sst{(6)}}}
\def\7{{\sst{(7)}}}
\def\8{{\sst{(8)}}}
\def\sst#1{{\scriptscriptstyle #1}}

\newcommand{\be}{\begin{equation}}
\newcommand{\ee}{\end{equation}}
\newcommand{\bea}{\setlength\arraycolsep{2pt} \begin{eqnarray}}
\newcommand{\eea}{\end{eqnarray}}
\newcommand{\nn}{\nonumber}

\begin{document}

\begin{flushright}
\hfill{ \
MIFPA-12-24\ \ \ \ }
\end{flushright}

\vspace{25pt}

\begin{center}

{\large \bf One-Loop Divergences in 6D Conformal Gravity}

{\large\bf}

\vspace{20pt}

Yi Pang

\vspace{10pt}

{\it George P. \& Cynthia Woods Mitchell  Institute
for Fundamental Physics and Astronomy,\\
Texas A\&M University, College Station, TX 77843, USA}

\vspace{40pt}

\underline{ABSTRACT}

\end{center}

\bigskip
Using Exact Renormalization Group Equation approach and background field method, we investigate the one-loop problem in a six-dimensional conformal gravity theory whose Lagrangian takes the same form as holographic Weyl anomaly of multiple coincident M5-branes. We choose the backgrounds to be the symmetric Einstein spaces including $S^6$, $CP^3$, $S^2\times S^4$, $S^2\times CP^2$, $S^3\times S^3$ and $S^2\times S^2\times S^2$. Evaluating the functional sums gives power-law and logarithmic divergences. We extract from the specific values of logarithmic divergence on above backgrounds, the coefficient in front of Euler density and two linear equations constraining the coefficients in front of three type-B conformal invariants. As a test of the effectiveness of Exact Renormalization Group Equation approach to quantum conformal gravity, we reexamine the one-loop problem in four-dimensional conformal gravity and confirm the logarithmic divergence derived from generalized Schwinger-DeWitt method.

\vspace{15pt}

\thispagestyle{empty}

\newpage

\tableofcontents

\newpage

\section{Introduction}
Conformal gravity is a class of gravity theory full of interests. It is constructed from Polynomials and conformally covariant derivatives of Weyl tensor. A Weyl transformation of the metric, $g_{\mu\nu}\rightarrow \Omega^2(x)g_{\mu\nu}$, is an exact symmetry of conformal gravity. Conformal gravity has appeared periodically in the literature for various reasons. In four dimensions, conformal gravity was considered as a possible UV completion of gravity in \cite{Zee:1980sj,Fradkin:1981iu,Adler:1982ri}.
It also appeared as counter term in $\mbox{AdS}_5$/$\mbox{CFT}_4$ computations \cite{hs}. Asymptotically AdS and Lifshitz solutions in four-dimensional conformal gravity have been studied in \cite{Riegert,mk,Klemm,Lu:2012xu}. The recent work \cite{Lu:2012xu}
reveals that these solutions can be related to AdS (dS)-Schwarzschild black holes in Einstein gravity by conformal transformations which also introduce new hair for asymptotically AdS black holes \cite{Lu:2012xu}. The principal reason that conformal gravity has not received general acceptance is because, since it is a higher-derivative theory, the propagating degrees of freedom contain ghost-like particles. Interestingly, it is suggested by \cite{mald} that in four-dimensional conformal gravity, the ghosts can be removed from the physical spectrum by imposing Neumann boundary condition which also selects Eintein solutions out of more numerous solutions of conformal graviy. Compared with Einstein gravity, conformal gravity possesses an advantage that the renormalized on-shell action of AdS-Schwarzschild black holes in Einstein gravity can be computed from the action of conformal gravity without referring to any regularization scheme. In \cite{Lu:2011ks}, this scenario has been generalized to a six-dimensional conformal gravity theory whose Lagrangian takes the same form as the holographic Weyl anomaly of multiple coincident M5-branes \cite{hs}.

The one-loop problem of four-dimensional conformal gravity has been tackled first in \cite{Fradkin:1981iu}, subsequently in \cite{Antoniadis:1992xu} and recently revisited by \cite{deBerredoPeixoto:2003pj} which adopted generalized Schwinger-DeWitt method.
One crucial point in the application of generalized Schwinger-DeWitt method is to
express the 4th order operator appearing in the gauge fixed quadratic action as the minimal form
\be
{\bf\Delta}^{(4)}=(\nabla^2)^2{\bf 1}+{\bf U}^{\mu\nu}\nabla_{\mu}\nabla_{\nu}+{\bf V}^{\mu}\nabla_{\mu}+{\bf W},
\ee
 by selecting a convenient gauge. With the aid of dimensional regularization, the one-loop effective action can be read from the $\textbf{b}_4$ (or $\textbf{a}_2$ ) coefficient in the heat kernel expansion of the 4th order minimal operator \cite{Barvinsky:1985an}.

We recall that in the study of quantum gravity, Exact Renormalization Group Equation (ERGE) is also a powerful tool which has been applied to explore the asymptotic safety in Einstein gravity \cite{Reuter:1996cp}, extended gravity with $R^2$ term \cite{Lauscher:2002sq,Codello:2006in} and topologically massive gravity \cite{Percacci:2010yk}. For a nice review on the application of ERGE in quantum gravity, the reader is referred to \cite{Codello:2008vh}. In this work, we will investigate the one-loop problem of the six-dimensional conformal gravity theory proposed in \cite{Lu:2011ks} by using Exact Renormalization Group Equation. In the calculation of one-loop divergences, we use background field method, therefore these divergences should be composed by terms invariant under general coordinate transformations. The backgrounds are chosen to be symmetric Einstein spaces including $S^6$, $CP^3$, $S^2\times S^4$, $S^2\times CP^2$, $S^3\times S^3$ and $S^2\times S^2\times S^2$. Since the spectra of differential operators
appearing in the gauge fixed quadratic action can be solved exactly for above backgrounds, the calculation of ERGE
is converted to the sums of eigenvalues restricted by a ``built-in'' cutoff. By computing these sums, we obtain the power-law and logarithmic divergences. Generically, the one-loop logarithmic divergence in a classically conformally invariant theory is also conformally invariant. Here we assume such property is retained in six-dimensional quantum conformal gravity. Therefore, the logarithmic divergence consists of six-dimensional Euler density $E_6$, three type-B conformal invariants $I_i (i=1,2,3)$ and total derivatives \cite{ds}. From the specific values of logarithmic divergence on above backgrounds, we extract the coefficient in front Euler density and two linear equations constraining the coefficients in front of three type-B conformal invariants. Before calculating the divergences in six-dimensional conformal gravity, we carry out a test of compatibility between ERGE approach and conformality of logarithmic divergence in quantum conformal gravity by applying ERGE approach to the one-loop problem in four-dimensional conformal gravity. It is shown that the results in \cite{Fradkin:1981iu,Antoniadis:1992xu,deBerredoPeixoto:2003pj} can be reproduced from ERGE approach. The power-law
divergences corresponding to quantum generations of cosmological constant, Einstein-Hilbert term and quadratic curvature terms are presented. In the framework of ERGE approach, these power-law divergences will depend on the explicit form of cutoff, however they appear to be qualitatively stable \cite{Codello:2008vh}. The appearance of non-conformally invariant terms seems to imply the breaking of conformality in conformal gravity at quantum level and the renormalizability would render the final theory to be a general cubic curvature theories. Therefore from the point of view of perturbatively renormalizability, a more accurate treatment will take into account the contribution of non-conformally invariant terms to ERGE. However, in the same spirit of \cite{'tHooft:1974bx} we will focus on the pure conformal gravity truncation in order to see the divergences solely from the quantum conformal gravity effects.

The rest of this paper is organized as follows. In section 2, we briefly review the essence of Exact Renormalization Group Equation. In section 3, we calculate the one-loop divergences in four-dimensional conformal gravity by evaluating the ERGE. In section 4, we apply the ERGE approach to a six-dimensional conformal gravity theory and obtain the one-loop divergences. Especially, we elaborate  on the derivation of logarithmic divergence. Conclusions are made in section 5.

\section{Preliminary of Exact Renormalization Equation }
The effective action describing physical phenomena at a typical momentum scale $k$ can be thought of as the result of integrating out all fluctuations of field with momenta larger than $k$. In this sense, $k$ can be regarded as the IR cutoff of the functional integration. The dependence of the effective action on $k$ is the Wilsonian RG flow. There are several ways of realizing this idea for practical use. One of the widely used approaches was proposed in \cite{Wetterich:1992yh}. According to \cite{Wetterich:1992yh}, a suppression term $\Delta S_{k}[\phi]$
\be\label{cutoff}
\Delta S_k[\phi]=\frac{1}{2}\int dx\phi(x)R_k({\cal O})\phi(x)
\ee
is added to the bare action $S[\phi]$ for the sake of suppressing the contribution from fluctuations with momentum lower than $k$.
The cutoff function $R_k(z)$ is required to be
a monotonically decreasing function in both $z$ and $k$. Due to the properties of $R_k(z)$, $k$
effectively plays the role of UV cutoff in the evaluation of the beta functions. One can define the $k$-dependent
generating functional of connected Green functions by
\be
e^{-W_k[J]}=\int D\phi\exp\biggl\{-S[\phi]-\Delta S_k[\phi]-\int dx J\phi\biggr\}
\ee
and a modified $k$-dependent Legendre transformation
\be
\Gamma_k[\phi]=W_k[J]-\int dx J\phi-\Delta S_k[\phi].
\ee
The functional $\Gamma_k$ is called the effective average action and satisfies
the following ``Exact Renormalization Group Equation''
\be\label{ERGE}
k\frac{d\Gamma_k}{dk}=\frac{1}{2}\mbox{Tr}[\Gamma^{(2)}_k+R_k]^{-1}k\frac{dR_k}{dk},
\ee
where $\Gamma^{(2)}_k$ is understood to be the exact connected two point function defined by
\be
\Gamma^{(2)}_k=\frac{\delta^2 \Gamma_k}{\delta\phi(x)\delta\phi(y)}.
\ee
Given a cutoff function $R_k(z)$, Eq.(\ref{ERGE}) describes the evolution of effective action along momentum scale $k$. The ERGE can also be applied to the theory which does not possess renormalizability and have infinitely many couplings as long as the system permits a useful truncation.

The ERGE can also been seen as a RG-improved evolution equation for one-loop effective action. To see this,
recall that given a bare action $S$, the one-loop effective action $\Gamma^{(1)}$ is
\be
\Gamma^{(1)}=S+\frac{1}{2}\mbox{Tr}\log\biggl[\frac{\delta^2S}{\delta\phi\delta\phi}\biggr].
\ee
 If we add to $S$ the cutoff term Eq.(\ref{cutoff}), the functional becomes
 \be
 \Gamma^{(1)}_{k}=S+\frac{1}{2}\mbox{Tr}\log\biggl[\frac{\delta^2S}{\delta\phi\delta\phi}+R_k\biggr].
 \ee
It satisfies the equation
 \be
k\frac{d\Gamma^{(1)}_k}{dk}=\frac{1}{2}\mbox{Tr}\biggl[\frac{\delta^2S}{\delta\phi\delta\phi}
+R_k\biggr]^{-1}k\frac{dR_k}{dk},
 \ee
 which takes a similar form as Eq.(\ref{ERGE}) except that in the r.h.s, the renormalized coupling constants
 are replaced by the bare ones. Therefore, the ERGE can be perceived as RG-improved effective action since the running of coupling constants is retained in the exact propagator. However, in one-loop calculation, usually it is good enough to consider a weakly improved ERGE by neglecting the derivatives of the coupling constants occurring in the r.h.s of ERGE \cite{Codello:2008vh}. This weakly improved ERGE will be adopted in later calculations.
\section{One-Loop Divergences in Four-Dimensional Conformal Gravity: A Test }
\subsection{Quadratic Gauge Fixed Action of Four-Dimensional Conformal Gravity}
In the following, we work throughout with a Euclidean metric
of positive signature. Four-dimensional conformal gravity is described by
\be
S=\int d^4x\sqrt{g}\biggl(\alpha C_{\mu\nu\sigma\rho}^2+\beta E_4\biggr),
\ee
where $C_{\mu\nu\sigma\rho}$ is the Weyl tensor and $E_4$ is the four-dimensional Euler density. Expanding the action around background metric $\bar{g}_{\mu\nu}$ as $g_{\mu\nu}=\bar{g}_{\mu\nu}+h_{\mu\nu}$,
In terms of powers of $h_{\mu\nu}$ we get
\be
S[g_{\mu\nu}]=\sum_n S^{(n)}[\bar{g}_{\mu\nu},h_{\mu\nu}].
\ee
Invariance of the original action under the conformal transformation
\be
g_{\mu\nu}\rightarrow e^{2\omega}g_{\mu\nu},
\ee
implies each $S^{(n)}$ should be invariant under the following transformations
\be
\bar{g}_{\mu\nu}\rightarrow e^{2\omega}\bar{g}_{\mu\nu},\quad h_{\mu\nu}\rightarrow e^{2\omega}h_{\mu\nu}.
\ee
Equivalently, to the leading order of $\omega$ we have
\be
\bar{g}_{\mu\nu}\frac{\delta S^{(n)}[\bar{g}_{\mu\nu},h_{\mu\nu}]}{\delta \bar{g}_{\mu\nu}}
+h_{\mu\nu}\frac{\delta S^{(n)}[\bar{g}_{\mu\nu},h_{\mu\nu}]}{\delta h_{\mu\nu}}=0.
\ee
On the other hand, the linearized conformal scaling symmetry
\be
\bar{g}_{\mu\nu}\rightarrow \bar{g}_{\mu\nu},\quad h_{\mu\nu}\rightarrow h_{\mu\nu}+2\omega \bar{g}_{\mu\nu},
\ee
states that $S^{(2)}$ is independent of $h\equiv g^{\mu\nu}h_{\mu\nu}$. Next, to fix the conformal invariance, we choose the following gauge fixing conditions for the
local conformal symmetry
\be
 h\equiv g^{\mu\nu}h_{\mu\nu}=0.
\ee
The gauge-fixing action for the diffeomorphism invariance is given by
\be
S_{GF}=\int d^4x\sqrt{g}\chi_{\mu}Y^{\mu\nu}\chi_{\nu},
\ee
where
\be
\chi_{\mu}=\nabla^{\mu}(h_{\mu\nu}-\ft14 g_{\mu\nu} h),\quad Y^{\mu\nu}=-\alpha(\Box g^{\mu\nu}-\ft{1}{3}\nabla^{\mu}\nabla^{\nu}).
\ee
Here we make a simplest choice for $Y^{\mu\nu}$, so that in $S^{(2)}+S_{GF}$ the 4th order operator sandwiched by $h_{\mu\nu}$ can be expressed as
standard minimal form.
Since $h$ does not enter $\chi^{\mu}$, it can be checked that the
ghost fields associated with fixing the local conformal invariance will not contribute to the one-loop divergences.
The Faddeev-Popov ghost action corresponding to $\chi_{\mu}$ is determined by
\be
S_{FP}=-\int d^4x\sqrt{g}\bar{C}_{\mu}\biggl(\Box g^{\mu\nu}+\ft12\nabla^{\mu}\nabla^{\nu}+R^{\mu\nu}\biggr)C_{\nu}.
\ee
Since $Y^{\mu\nu}$ contains differential operators, it will contribute $-\ft12\ln\mbox{det}Y^{\mu}_{~~\nu}$ to the one-loop effective action. Exponentiating this logarithmic determinant, we obtain the action for Nielsen-Kallosh ghosts as follows
\be
S_{NK}=\int d^4x\sqrt{g}\bar{\omega}_{\mu}Y^{\mu\nu}\omega_{\nu}+\int d^4x\sqrt{g}f_{\mu}Y^{\mu\nu}f_{\nu}.
\ee
In above expressions, $\bar{\omega}_{\mu}$ and $\omega_{\mu}$ are complex anticommuting vectors, $f_{\mu}$
is a real commuting vector.

In order to achieve diagonalization of the inverse propagator we decompose the fluctuation $h_{\mu\nu}$ into
irreducible parts
\be
h_{\mu\nu}=h^{TT}_{\mu\nu}+\bar{\nabla}_{\mu}\xi^{T}_{\nu}
+\bar{\nabla}_{\nu}\xi^{T}_{\mu}+\bar{\nabla}_{\mu}\bar{\nabla}_{\nu}\sigma
-\ft14\bar{g}_{\mu\nu}\bar{\Box}\sigma+\ft14\bar{g}_{\mu\nu}h,\label{decompositon}
\ee
where $\xi^T_{\lambda}$ satisfies $\bar{\nabla}^{\lambda}\xi^T_{\lambda}=0$ and $h^{TT}_{\mu\nu}$ satisfies
$\bar{g}^{\mu\nu}h^{TT}_{\mu\nu}=0$ and $\bar{\nabla}^{\mu}h^{TT}_{\mu\nu}=0$. To achieve the maximal factorization of higher order operators, we choose the backgrounds to be symmetric Einstein spaces, for which
\be
\bar{\nabla}_{\lambda}\bar{R}_{\mu\nu\sigma\rho}=0;\quad\bar{ R}_{\mu\nu}=\ft{\bar{R}}{4}\bar{g}_{\mu\nu}.
\ee
It is noted that in passing from a path integral over field $h_{\mu\nu}$ to one over the fields $(h^{TT}_{\mu\nu}, \xi^{T}_{\nu},\sigma,h )$, one has to take into account the appropriate Jacobian factors. These Jacobian will be
exactly canceled by other Jacobians arising from suitable redefinitions of $\xi^{T}_{\nu}$ and $\sigma$ (see below). In the following, we drop the ``bar'' for notational simplicity.
Inserting Eq.(\ref{decompositon}) into $S^{(2)}+S_{GF}$, we find

\bea\label{4D quadratic action}
S^{(2)}+S_{GF}&=&-\frac{\alpha}{2}\int  d^4x\sqrt{g}\biggl(
h^{TT\mu\nu}(\Delta_{L}-\ft{R}{3})(\Delta_{L}-\ft{R}{2})h^{ TT}_{\mu\nu}\biggr)\nn\\
&&-\alpha\int d^4x\sqrt{g}\biggl({\xi}^T_{\mu}\Box(\Box+\ft{R}{4})^2{\xi}^{T\mu}
-\ft38{\sigma}\Box(\Box+\ft{R}{3})^2(\Box+\ft{3R}{8}){\sigma}\biggr),
\eea
where $\Delta_{L}$ is the Lichnerowicz operator acting on $h_{\mu\nu}$. On a $d$-dimensional Einstein manifold with $R_{\mu\nu}=\ft{R}{d} g_{\mu\nu}$, the Lichnerowicz operator is defined by
\be
\Delta_{L}h_{\mu\nu}=-{\Box} h_{\mu\nu}-2{R}_{\mu\sigma\nu\rho}h^{\sigma\rho}+\ft{2R}{d} h_{\mu\nu}.
\ee
 Because any Einstein metric solves the equaitons of motion of four-dimensional conformal gravity, the one-loop calculation performed in this paper is ``on-shell". Similarly, we decompose the ghost fields as
\be\label{deghost}
C_{\mu}=C_{\mu}^{T}+\partial_{\mu}C;\quad \omega_{\mu}=\omega_{\mu}^{T}+\partial_{\mu}\omega;\quad
f_{\mu}=f_{\mu}^{T}+\partial_{\mu}f,
\ee
and similarly for $\bar{C}_{\mu}$ and $\bar{\omega}_{\mu}$. This leads to
\bea\label{ghostaction}
S_{FP}&=&\int d^4x\sqrt{g}\biggl(\bar{C}^{T\mu}(-\Box-\ft{R}{4})C^{T}_{\mu}+\ft32\bar{C}\Box(\Box+\ft{R}{3})C\biggr),\nn\\
S_{NK}&=&-\alpha\int d^4x\biggl(\bar{\omega}^{T\mu}\Box\omega ^T_{\mu}-\ft{2}{3}\bar{\omega}\Box(\Box+\ft{3R}{8})\omega\biggr)\nn\\
&&-\alpha\int d^4x\biggl(f^{T\mu}\Box f^T_{\mu}-\ft{2}{3}f\Box(\Box+\ft{3R}{8})f\biggr).
\eea

\subsection{The Cutoff}
In this section, we define the cutoff and express the ERGE as functional traces.
We make the field redefinitions
\be
\hat{\xi}^T_{\mu}=\sqrt{(-\Box-\ft{R}{4})}\xi^T_{\mu};\quad\hat{\sigma}=\sqrt{-\Box(-\Box-\ft{R}{3})}\sigma;
\ee
\be
\hat{C}=\sqrt{-\Box}C;\quad \hat{\omega}=\sqrt{-\Box}\omega;\quad\hat{f}=\sqrt{-\Box}f,
\ee
whose Jacobian factors cancel those coming from Eq.(\ref{decompositon}) and Eq.(\ref{deghost}) \cite{Fradkin2,Dou:1997fg}. Then the action Eq.(\ref{4D quadratic action}) becomes
\be
S^{(2)}+S_{GF}=-\frac{\alpha}{2}\int  d^4x\sqrt{g}\biggl[
h^{TT\mu\nu}\Delta_{h\mu\nu}^{~~~\rho\sigma}h^{ TT}_{\rho\sigma}+q_1\hat{{\xi}}^{T\mu}\Delta_{\xi\mu}^{~~\nu}\hat{{\xi}}^T_{\nu}+q_2\hat{\sigma}\Delta_{\sigma}\hat{\sigma}\biggr],
\ee
where we have defined the operators
\bea\label{4dhopertators}
&&\Delta_{h\mu\nu}^{~~\rho\sigma}=\Delta_{h,1\mu\nu}^{~~~~\lambda\delta}\Delta_{h,2\lambda\delta}^{~~~~\rho\sigma},\quad
\Delta_{h,1\mu\nu}^{~~~~\rho\sigma}=(\Delta_{L}-\ft{R}{3})\delta^{(\rho}_{(\mu}\delta^{~\sigma)}_{\nu)},\quad
\Delta_{h,2\mu\nu}^{~~~~\rho\sigma}=(\Delta_{L}-\ft{R}{2})\delta^{(\rho}_{(\mu}\delta^{~\sigma)}_{\nu)},\nn\\
&&\Delta_{\xi\mu}^{~~\nu}=\Delta_{\xi,1\mu}^{~~~\lambda}\Delta_{\xi,2\lambda}^{~~~\nu},\quad
\Delta_{\xi,1\mu}^{~~~\nu}=\Box\delta_{\mu}^{\nu},\quad
\Delta_{\xi,2\mu}^{~~~\nu}=(-\Box-\ft{R}{4})\delta_{\mu}^{\nu},\nn\\
&&\Delta_{\sigma}=\Delta_{\sigma1}\Delta_{\sigma2},\quad
\Delta_{\sigma1}=(-\Box-\ft{R}{3}),\quad
\Delta_{\sigma2}=(-\Box-\ft{3R}{8}),
\eea
and coefficients
\be
q_1=2;\quad q_2=-\frac{3}{4}.
\ee
Similarly, the ghost actions Eq.(\ref{ghostaction}) become
\bea\label{ghostaction}
S_{FP}&=&\int d^4x\sqrt{g}\biggl(\bar{C}^{T\mu}\Delta^{~~~~\nu}_{C^T\mu}C_{T\nu}+q_3\hat{\bar{C}}\Delta_C\hat{C}\biggr),\nn\\
S_{NK}&=&\alpha\int d^4x\biggl(\bar{\omega}^{T\mu}\Delta^{~~~~\nu}_{\omega^T\mu}\omega ^T_{\nu}+q_4\hat{\bar{\omega}}\Delta_C\hat{\omega}
+f^{T\mu}\Delta^{~~~~\nu}_{f^T\mu} f^T_{\nu}+q_5\hat{f}\Delta_f\hat{f}\biggr)
\eea
where we defined operators
\bea
&&\Delta^{~~~~\nu}_{C^T\mu}=(-\Box-\ft{R}{4})\delta_{\mu}^{~\nu},\quad\Delta_C=(-\Box-\ft{R}{3})\nn\\
&&\Delta^{~~~~\nu}_{\omega^T\mu}=-\Box\delta_{\mu}^{~\nu},\quad \Delta_\omega=(-\Box-\ft{3R}{8})\nn,\\
&&\Delta^{~~~~\nu}_{f^T\mu}=-\Box\delta_{\mu}^{~\nu},\quad \Delta_f=(-\Box-\ft{3R}{8})
\eea
and coefficients
\be
q_3=\ft32,\quad q_4=\ft{2}{3},\quad q_5=\ft{2}{3}.
\ee
We define the gauge fixed inverse propagator
\be
{\cal O}=-\frac{\alpha}{2} \left(
  \begin{array}{ccc}
    \Delta_{h} &  &  \\
& \Delta_{\xi} &  \\
     &  & \Delta_{\sigma} \\
  \end{array}
\right).
\ee
For each spin component of $h_{\mu\nu}$, we choose the cutoff to have the following forms
\be
{\cal R}_k=-\frac{\alpha}{2} \left(
  \begin{array}{ccc}
   R_k( \Delta_{h}) &  &  \\
&  R_k(\Delta_{\xi}) &  \\
     &  & R_k( \Delta_{\sigma}) \\
  \end{array}
\right),
\ee
where $R_k=(k^4-z)\theta(k^4-z)$. Discussions of cutoffs for ghost fields is similar, however since the corresponding operators are second order, $R_k=(k^2-z)\theta(k^2-z)$. With above preparations, employing the ERGE, we can express the evolution equation of effective action with respect to the renormalization group ``time $t\equiv \ln k$'' as
\bea\label{4dsum}
\partial_t\Gamma_k&=&2\sum_nd_n^{h^{TT}}\theta(k^4-\lambda^{h^{TT}}_n)
+2\sum_n'd_n^{\xi^{T}}\theta(k^4
-\lambda^{\xi^{T}}_n)+2\sum''_nd_n^{\sigma}\theta(k^4-\lambda^{\sigma}_n)
\nn\\&&-2\sum_nd_n^{C^T}\theta(k^2-\lambda^{C^T}_n)-2\sum'_nd_n^{C}\theta(k^2-\lambda^{C}_n)-2\sum_nd_n^{\omega^T}\theta(k^2-\lambda^{\omega^T}_n),\nn\\
&&-2\sum'_nd_n^{\omega}\theta(k^2-\lambda^{\omega}_n)+\sum_nd_n^{f^T}\theta(k^2-\lambda^{f^T}_n)+\sum'_nd_n^{f}\theta(k^2-\lambda^{f}_n),
\eea
where $\lambda_n$ and $d_n$ stand for the level-$n$ eigenvalue and corresponding degeneracy respectively.
Note that single ``prime" for $\xi$ states that the vector modes corresponding to the Killing vectors are excluded.
Single ``prime" for scalars denotes that the constant scalar modes should not be taken into account. Double ``prime" for $\sigma$ means both the constant modes and the modes corresponding to conformal Killing vectors are not taken into account.
\subsection{Evaluation of Functional Traces}

By dimensional analysis, the evolution equation should have the following form
\be
\partial_t\Gamma_k=\biggl({\cal A}_0(k)R^{-2}+{\cal A}_2(k)R^{-1}+{\cal A}_4+{\cal O}(R)\biggr),
 \ee
where $R$ is the scalar curvature of background metric.
Utilizing the spectra of Hodge-de Rahm and  Lichnerowicz operators on $S^4$, $CP^2$ (see Tables IV, V in Appendices A, B) and $S^2\times S^2$\footnote{Eigenfunctions of Hodge-de Rahm and Lichnerowicz on $S^2\times S^2$ can be constructed from the bilinear of Harmonics on $S^2$. For the recent application of harmonics on $S^2$, the reader is referred to \cite{Pang:2012xs}. }, We obtain all the divergences in the evolution equation of one-loop effective action. These results are exhibited in Table 1. In the calculation, it is assumed that $k^2\gg R$.
\begin{table}[ht]
\centering
\begin{tabular}{|c|c|c|c|}
  \hline
 Manifold &${\cal A}_0(k)$&${\cal A}_2(k)$&${\cal A}_4$ \\
  \hline\hline
 $S^4$   &$72k^4$  & $-136 k^2 $   &  -$\frac{87}{5}$        \\
  \hline
  $CP^2$  &$54k^4$   & $ -102k^2$   & -$\frac{31}{5}$    \\
  \hline
  $S^2\times S^2$ &$48k^4$      & $-\frac{272k^2 }{3}$  & $\frac{26}{45}$  \\
  \hline
\end{tabular}
\caption{In this table, we present the divergences appearing in $\partial_t\Gamma_k$. }
\label{Table1}
\end{table}
Upon integrating $\partial_t\Gamma_k$ to certain UV cutoff $\Lambda$, one can obtain the power-law and logarithmic
divergences in one-loop effective action. Among these divergences, the most interesting one is the logarithmic divergence which can be parameterized as
\be\label{4dlog}
\Gamma_{\mbox{log}\Lambda}=\frac{\log\Lambda}{(4\pi)^2}\int d^4x\sqrt{g}\biggl(a E_4+ cC^2_{\mu\nu\sigma\rho}+d\nabla_{\mu}J^{\mu}\biggr),
\ee
\begin{table}[ht]
\centering
\begin{tabular}{|c|c|c|c|}
  \hline
 Manifold &Vol/$(4\pi)^2$   & $C_{\mu\nu\sigma\rho}^2$&$E_4$ \\
  \hline\hline
 $S^4$   &$\frac{24}{R^2}$     &   0&$\frac{R^2}{6}$       \\
  \hline
  $CP^2$  &$\frac{18}{R^2}$     & $ \frac{R^2}{6}$   &$\frac{R^2}{3}$      \\
  \hline
  $S^2\times S^2$       & $\frac{16}{R^2}$    & $\frac{R^2}{3}$  & $\frac{R^2}{2}$   \\
  \hline
\end{tabular}
\caption{In this table, we present the values of Euler density and Weyl squared on $S^4$, $CP^2$ and $S^2\times S^2$.}
\label{Table2}
\end{table}

When evaluated on $S^4$, $CP^2$ and $S^2\times S^2$, Eq.(\ref{4dlog}) should reproduce the corresponding ${\cal A}_4$ coefficients. Since $\nabla_{\mu}J^{\mu}$ vanishes on Einstein spaces, using results of Table 2, we find
\be
a=-\frac{87}{20},\quad c=\frac{199}{30}.
\ee
These two coefficients agree with those obtained in \cite{Fradkin:1981iu,Antoniadis:1992xu,deBerredoPeixoto:2003pj}
after a Wick rotation from Euclidean signature to Lorentzian.

Before proceeding to six dimensions, we would like to discuss another way
of calculating the one-loop divergences in four-dimensional conformal gravity
based on heat kernel expansion. By a straightforward calculation, we notice that
on $S^4$, $CP^2$ and $S^2\times S^2$
the functional sums in Eq.(\ref{4dsum}) is equivalent to the following expression up to ${\cal O}(k^{-2})$
\bea\label{4dresum}
\partial_t\Gamma_k&=&\sum_nd_n^{h^{TT}}\theta(k^2-\lambda^{h^{TT}}_{n,1})
+\sum_n'd_n^{\xi^{T}}\theta(k^2
-\lambda^{\xi^{T}}_{n,1})+\sum''_nd_n^{\sigma}\theta(k^2-\lambda^{\sigma}_{n,1})
\nn\\
&&+\sum_nd_n^{h^{TT}}\theta(k^2-\lambda^{h^{TT}}_{n,2})
+\sum_n'd_n^{\xi^{T}}\theta(k^2
-\lambda^{\xi^{T}}_{n,2})+\sum''_nd_n^{\sigma}\theta(k^2-\lambda^{\sigma}_{n,2})
\nn\\
&&-2\sum_nd_n^{C^T}\theta(k^2-\lambda^{C^T}_n)-2\sum'_nd_n^{C}\theta(k^2-\lambda^{C}_n)-2\sum_nd_n^{\omega^T}\theta(k^2-\lambda^{\omega^T}_n),\nn\\
&&-2\sum'_nd_n^{\omega}\theta(k^2-\lambda^{\omega}_n)+\sum_nd_n^{f^T}\theta(k^2-\lambda^{f^T}_n)+\sum'_nd_n^{f}\theta(k^2-\lambda^{f}_n)+ {\cal O}(k^{-2}),
\eea
where each sum over eigenvalues of a 4th order operator has been splitted into two
sums over eigenvalues of two 2nd order operators which factor the 4th order operator (see Eq.(\ref{4dhopertators})).

 Generically, for a $d$-dimensional 2nd order operator $\Delta$
\be\label{heatexp}
\mbox{Tr}W(\Delta)=\biggl(Q_{\frac{d}{2}}(W)\textbf{B}_0(\Delta)+
Q_{\frac{d}{2}-1}(W)\textbf{B}_2(\Delta)+\ldots+Q_0(W)\textbf{B}_d(\Delta)+\ldots\biggr),
\ee
where $W(\Delta)$ is a function of $\Delta$, $\textbf{B}_n$ is related to the Seeley-DeWitt coefficient
${\textbf b}_n$ by
\be
{\textbf B}_n=\frac{1}{(4\pi)^{\frac{d}{2}}}\int d^dx\sqrt{g}{\textbf b}_n,
 \ee
and based on $\tilde{W}$, the Laplace anti-tranform of $W$, $Q_n$ is defined by
\be
Q_n(W)=\int_0^{\infty}dss^{-n}\tilde{W}(s).
\ee
Applying Eq.(\ref{heatexp}) to Eq.(\ref{4dresum}), we otain
\bea
&&\partial_t\Gamma_k=\mbox{Tr}W(\Sigma_i\gamma_i\Delta_i)\nn\\
&&=\biggl(Q_{\frac{d}{2}}(W){\textbf B}_0^{\mbox{t}}+
Q_{\frac{d}{2}-1}(W){\textbf B}_2^{\mbox{t}}+\ldots+Q_0(W){\textbf B}_d^{\mbox{t}}+\ldots\biggr),
\quad {\textbf B}_n^{\mbox{t}}=\Sigma_i\gamma_i{\textbf B}_n(\Delta_i),
\eea
where $W(z)=\theta(k^2-z)$, $\Delta_i$s are 2nd order operators whose eigenvalues appearing in the sums Eq.(\ref{4dresum}) and
coefficients $\gamma_i$s can be read from Eq.(\ref{4dresum}) directly. On
symmetric Einstein spaces, we find
\bea\label{heatkernel}
 &&{\cal A}_0=\frac{1}{(4\pi)^2}\int d^dx\sqrt{g}{\textbf b}_0^{\mbox{t}}Q_2,\quad{\textbf b}_0^{\mbox{t}}=6,\quad Q_2=\frac{1}{2}k^4,\nn\\
&&{\cal A}_2=\frac{1}{(4\pi)^2}\int d^dx\sqrt{g}{\textbf b}_2^{\mbox{t}} Q_1,\quad{\textbf b}_2^{\mbox{t}}=-\frac{17R}{3},\quad \quad Q_1=k^2;\nn\\
&&{\cal A}_4=\frac{1}{(4\pi)^2}\int d^dx\sqrt{g}{\textbf b}_4^{\mbox{t}}Q_0,\quad{\textbf b}_4^{\mbox{t}}=\frac{137}{60}R_{\mu\nu\sigma\rho}^2-\frac{1}{30}R_{\mu\nu}^2-\frac{79}{72}R^2,\quad Q_0=1.
\eea
In the derivation of above formulae, the property of symmetric Einstein spaces has been used. Inserting the specific metrics
on $S^4$, $CP^2$ and $S^2\times S^2$ to above formulae, one can immediately reproduce the results listed in Table 1. Since ${\textbf b}_0^{\mbox{t}}$ is equivalent to the number of degrees of freedom in a theory, thus we see that four-dimensional conformal gravity contains 6 degrees of freedom.
\section{One-Loop Divergences in Six-Dimensional Conformal Gravity }
\subsection{Quadratic Gauge Fixed Action of Six-Dimensional Conformal Gravity}
In six dimensions, there are three type-B conformal invariants consisting of\footnote{Here we following the notations of \cite{bcn}} polynomials and covariant derivatives of Weyl tensor
\bea
I_1 &=& C_{\mu\rho\sigma\nu} C^{\mu\alpha\beta\nu}
 C_{\alpha}{}^{\rho\sigma}{}_\beta\,,\nn\\
I_2 &=& C_{\mu\nu\rho\sigma} C^{\rho\sigma\alpha\beta}
           C_{\alpha\beta}{}^{\mu\nu}\,,\nn\\
I_3 &=& C_{\mu\rho\sigma\lambda}\Big(\delta^\mu_\nu\, \Box +
    4R^\mu{}_\nu - \fft65 R\, \delta^\mu_\nu\Big) C^{\nu\rho\sigma\lambda}
   + \nabla_\mu J^\mu\,,
\eea
where $\nabla_\mu J^\mu$ vanishes on symmetric Einstein spaces.
In general, a Lagrangian constructed from general linear combinations of $I_i (i=1,2,3)$  will give equations of motion that are not satisfied by arbitrary Einstein metrics. However, for a
specific choice of the combination coefficients,
the equations of motion will be satisfied by any Einstein
metric. This same linear combination has the feature that,
modulo total derivatives, all terms of cubic and quadratic
order in the Riemann tensor are absent \cite{Lu:2011ks}. Then we achieve the following Lagrangian
taking the same form as holographic Weyl anomaly of multiple coincident M5-branes.

\be
e^{-1} {\cal L}_{\rm conf}^{6D} = \alpha( 4 I_1 + I_2 - \ft13 I_3 -
\ft1{24} E_6 ).
\ee
Note that $E_6$ is the Euler
density defined by
\begin{equation}
E_6= \epsilon_{\mu_1\nu_1 \mu_2\nu_2 \mu_3\nu_3} \epsilon^{\rho_1
\sigma_1 \rho_2 \sigma_2 \rho_3\sigma_3} R^{\mu_1
\nu_1}{}_{\rho_1\sigma_1} R^{\mu_2 \nu_2}{}_{\rho_2\sigma_2}
R^{\mu_3 \nu_3}{}_{\rho_3\sigma_3}\,.\label{d6conflag}
\end{equation}
After some algebraic manipulations the above Lagrangian can be recast into
\be
e^{-1} {\cal L}_{\rm conf}^{6D}=\alpha
\Big(RR^{\mu\nu}R_{\mu\nu}-\ft{3}{25} R^3-2R^{\mu\nu}R^{\rho\sigma}
R_{\mu\rho\nu\sigma} -R^{\mu\nu}\Box R_{\mu\nu}+\ft{3}{10}R\Box
R\Big)+  t.d\,.
\ee

Expanding above action around background metric $\bar{g}_{\mu\nu}$ as $g_{\mu\nu}=\bar{g}_{\mu\nu}+h_{\mu\nu}$ and
using the six-dimensional field decomposition
\be\label{6ddecompose}
h_{\mu\nu}=h^{TT}_{\mu\nu}+\bar{\nabla}_{\mu}\xi^{T}_{\nu}
+\bar{\nabla}_{\nu}\xi^{T}_{\mu}+\bar{\nabla}_{\mu}\bar{\nabla}_{\nu}\sigma
-\ft16\bar{g}_{\mu\nu}\bar{\Box}\sigma+\ft16\bar{g}_{\mu\nu}h,
\ee
we find that to the quadratic order in $h_{\mu\nu}$ the action of six-dimensional conformal gravity is given by
 \be\label{6dhaction}
S^{(2)}=-\frac{\alpha}{2}\int d^6x\sqrt{g}\biggl(h^{ TT\mu\nu}(\Delta_{L}-\ft{2\bar{R}}{15})
(\Delta_{L}-\ft{\bar{R}}{5})(\Delta_{L}-\ft{\bar{R}}{3})h^{TT}_{\mu\nu}\biggr),
\ee
where $\xi^{T}_{\nu},\sigma,h$ disappear, because they are gauge degrees of freedom in a conformally invariant
gravity theory and $S^{(2)}$ is gauge independent. The background is understood to be a symmetric Einstein space with
\be
\bar{\nabla}_{\lambda}\bar{R}_{\mu\nu\sigma\rho}=0;\quad\bar{ R}_{\mu\nu}=\ft{\bar{R}}{6}\bar{g}_{\mu\nu}.
\ee
In the following, ``bar" is removed from the notation of background quantities for simplicity.
Similar to the four-dimensional case, Jacobian factors coming from replacing $h_{\mu\nu}$ by $(h^{TT}_{\mu\nu},
\xi^{T}_{\nu},\sigma,h )$ will be exactly canceled by other Jacobians arising from suitable redefinitions of $\xi^{T}_{\nu}$ and $\sigma$ (see below).
To fix the conformal invariance, we choose the following gauge fixing conditions for the
local conformal symmetry
\be
 h\equiv g^{\mu\nu}h_{\mu\nu}=0.
\ee
In order that the 6th order operator sandwiched by $h_{\mu\nu}$ takes the standard minimal form
\be
{\bf\Delta}^{(6)}=(\nabla^2)^3{\bf 1}+{\bf U}^{\mu\nu\rho\sigma}\nabla_{\mu}\nabla_{\nu}\nabla_{\rho}\nabla_{\sigma}
+{\bf V}^{\mu\nu\rho}\nabla_{\mu}\nabla_{\nu}\nabla_{\rho}+{\bf W}^{\mu\nu}\nabla_{\mu}\nabla_{\nu}
+{\bf X}^{\mu}\nabla_{\mu}+{\bf Z},
\ee
the gauge-fixing action for the diffeomorphism invariance is chosen to be
\be
S_{GF}=\int d^4x\sqrt{g}\chi_{\mu}Y^{\mu\nu}\chi_{\nu},
\ee
where
\be
\chi_{\mu}=\nabla^{\mu}(h_{\mu\nu}-\ft16 g_{\mu\nu} h),\quad Y^{\mu\nu}=\ft{\alpha}{2}(\Box^2 g^{\mu\nu}-\ft{2}{5}\nabla^{\mu}\Box\nabla^{\nu}).
\ee
Inserting the decomposition Eq.(\ref{6ddecompose}) into the gauge-fixing action leads to
\be\label{6dGF}
S_{GF}=\frac{\alpha}{2}\int d^6x\sqrt{g}\biggl(\xi^{T\mu}\Box^2(\Box+\ft{R}{6})^2\xi^ T_{\mu}-\ft{5}{12}\sigma\Box(\Box+\ft{R}{5})^2(\Box^2+\ft{5R}{9}\Box+\ft{5R^2}{108})\sigma\biggr).
\ee
Since $h$ does not enter $\chi^{\mu}$, it can be checked that the conformal
ghost fields associated with fixing the local conformal invariance will not contribute to the one-loop divergences.
The Faddeev-Popov ghost action corresponding to $\chi_{\mu}$ is determined by
\be
S_{FP}=-\int d^6x\sqrt{g}\bar{C}_{\mu}\biggl(\Box g^{\mu\nu}+\ft23\nabla^{\mu}\nabla^{\nu}+R^{\mu\nu}\biggr)C_{\nu}.
\ee
Because $Y^{\mu\nu}$ contains differential operators, it will contribute $-\ft12\ln\mbox{det}Y^{\mu}_{~~\nu}$ to the one-loop effective action. Exponentiating this logarithmic determinant, we obtain the action for Nielsen-Kallosh ghosts as follows
\be
S_{NK}=\int d^6x\sqrt{g}\bar{\omega}_{\mu}Y^{\mu\nu}\omega_{\nu}+\int d^6x\sqrt{g}f_{\mu}Y^{\mu\nu}f_{\nu}.
\ee
In above expressions, $\bar{\omega}_{\mu}$ and $\omega_{\mu}$ are complex anticommuting vectors, $f_{\mu}$
is a real commuting vector. We decompose the ghost fields as
\be\label{6ddeghost}
C_{\mu}=C_{\mu}^{T}+\partial_{\mu}C;\quad \omega_{\mu}=\omega_{\mu}^{T}+\partial_{\mu}\omega;\quad
f_{\mu}=f_{\mu}^{T}+\partial_{\mu}f,
\ee
and similarly for $\bar{C}_{\mu}$ and $\bar{\omega}_{\mu}$. This leads to
\bea\label{6dghostaction}
S_{FP}&=&\int d^6x\sqrt{g}\biggl(\bar{C}^{T\mu}(-\Box-\ft{R}{6})C^{T}_{\mu}+\ft53\bar{C}\Box(\Box+\ft{R}{5})C\biggr),\nn\\
S_{NK}&=&\frac{\alpha}{2}\int d^6x\biggl(\bar{\omega}^{T\mu}\Box^2\omega^ T_{\mu}-\ft{3}{5}\bar{\omega}\Box(\Box^2+\ft{5R}{9}\Box+\ft{5R^2}{108})\omega\biggr)\nn\\
&&+\frac{\alpha}{2}\int d^6x\biggl(f^{T\mu}\Box^2 f^ T_{\mu}-\ft{3}{5}f\Box(\Box^2+\ft{5R}{9}\Box+\ft{5R^2}{108})f\biggr).
\eea

\subsection{The Cutoff}
In this section, we define the cutoff and express the ERGE as functional traces.
We employ the field redefinitions
\be
\hat{\xi}^T_{\mu}=\sqrt{(-\Box-\ft{R}{6})}\xi^T_{\mu};\quad\hat{\sigma}=\sqrt{-\Box(-\Box-\ft{R}{5})}\sigma;
\ee
\be
\hat{C}=\sqrt{-\Box}C;\quad \hat{\omega}=\sqrt{-\Box}\omega;\quad\hat{f}=\sqrt{-\Box}f,
\ee
whose Jacobian factors cancel those coming from Eq.(\ref{6ddecompose}) and Eq.(\ref{6ddeghost}).
Then the actions Eq.(\ref{6dhaction}) and Eq.(\ref{6dGF}) become
\be
S^{(2)}+S_{GF}=-\frac{\alpha}{2}\int  d^6x\sqrt{g}\biggl[
h^{TT\mu\nu}\Delta_{h\mu\nu}^{~~~\rho\sigma}h^{ TT}_{\rho\sigma}+p_1\hat{{\xi}}^{T\mu}\Delta_{\xi\mu}^{~~\nu}\hat{{\xi}}^T_{\nu}+p_2\hat{\sigma}\Delta_{\sigma}\hat{\sigma}\biggr],
\ee
where we have defined the operators
\bea\label{hopertators}
&&\Delta_{h\mu\nu}^{~~\rho\sigma}=\Delta_{h,1\mu\nu}^{~~~~\lambda\delta}\Delta_{h,2\lambda\delta}^{~~~~\alpha\beta}\Delta_{h,3\alpha\beta}^{~~~~\rho\sigma}\nn\\
&&\Delta_{h,1\mu\nu}^{~~~~\rho\sigma}=(\Delta_{L}-\ft{2R}{15})\delta^{(\rho}_{(\mu}\delta^{~\sigma)}_{\nu)},\quad
\Delta_{h,2\mu\nu}^{~~~~\rho\sigma}=(\Delta_{L}-\ft{R}{5})\delta^{(\rho}_{(\mu}\delta^{~\sigma)}_{\nu)},\quad
\Delta_{h,3\mu\nu}^{~~~~\rho\sigma}=(\Delta_{L}-\ft{R}{3})\delta^{(\rho}_{(\mu}\delta^{~\sigma)}_{\nu)},\nn\\
&&\Delta_{\xi\mu}^{~~\nu}=\Delta_{\xi,1\mu}^{~~~\lambda}\Delta_{\xi,2\lambda}^{~~~\sigma}\Delta_{\xi,3\sigma}^{~~~\nu},\nn\\
&&\Delta_{\xi,1\mu}^{~~~\nu}=\Delta_{\xi,2\mu}^{~~~\nu}=-\Box\delta_{\mu}^{\nu},\quad
\Delta_{\xi,3\mu}^{~~~\nu}=(-\Box-\ft{R}{6})\delta_{\mu}^{\nu},\nn\\
&&\Delta_{\sigma}=\Delta_{\sigma1}\Delta_{\sigma2},\nn\\
&&\Delta_{\sigma1}=(-\Box-\ft{R}{5}),\quad
\Delta_{\sigma2}=(\Box^2+\ft{5R}{9}\Box+\ft{5R^2}{108}),
\eea
and coefficients
\be
p_1=-1;\quad p_2=-\ft{5}{12}.
\ee
Similarly, the ghost actions Eq.(\ref{6dghostaction}) become
\bea
S_{FP}&=&\int d^6x\sqrt{g}\biggl(\bar{C}^{T\mu}\Delta^{~~~~\nu}_{C^T\mu}C^{T}_{\nu}+p_3\hat{\bar{C}}\Delta_C\hat{C}\biggr),\nn\\
S_{NK}&=&\frac{\alpha}{2}\int d^6x\biggl(\bar{\omega}^{T\mu}\Delta^{~~~~\nu}_{\omega^T\mu}\omega^ T_{\nu}+p_4\hat{\bar{\omega}}\Delta_\omega\hat{\omega}\biggr)\nn\\
&&+\frac{\alpha}{2}\int d^6x\biggl(f^{T\mu}\Delta^{~~~~\nu}_{f^T\mu}f^ T_{\nu}+p_5\hat{f}\Delta_f\hat{f}\biggr),
\eea
where we defined operators
\bea
&&\Delta^{~~~~\nu}_{C^T\mu}=(-\Box-\ft{R}{6})\delta_{\mu}^{~\nu},\quad\Delta_C=(-\Box-\ft{R}{5})\nn\\
&&\Delta^{~~~~\nu}_{\omega^T\mu}=\Box^2\delta_{\mu}^{~\nu},\quad \Delta_\omega=(\Box^2+\ft{5R}{9}\Box+\ft{5R^2}{108})\nn,\\
&&\Delta^{~~~~\nu}_{f^T\mu}=\Box^2\delta_{\mu}^{~\nu},\quad \Delta_f=(\Box^2+\ft{5R}{9}\Box+\ft{5R^2}{108})
\eea
and coefficients
\be
p_3=\ft53,\quad p_4=\ft{3}{5},\quad p_5=\ft{3}{5}.
\ee
We define the gauge fixed inverse propagator
\be
{\cal O}=-\frac{\alpha}{2} \left(
  \begin{array}{ccc}
    \Delta_{h} &  &  \\
& \Delta_{\xi} &  \\
     &  & \Delta_{\sigma} \\
  \end{array}
\right).
\ee
For each spin component of $h_{\mu\nu}$, we choose the cutoff to have the following forms
\be
{\cal R}_k=-\frac{\alpha}{2} \left(
  \begin{array}{ccc}
   R_k( \Delta_{h}) &  &  \\
&  R_k(\Delta_{\xi}) &  \\
     &  & R_k( \Delta_{\sigma}) \\
  \end{array}
\right),
\ee
where $R_k=(k^6-z)\theta(k^6-z)$. Discussions of cutoffs for ghost fields is similar, however we choose
$R_k=(k^2-z)\theta(k^2-z)$ for Faddeev-Popov ghost and $R_k=(k^4-z)\theta(k^4-z)$ for Nielsen-Kallosh ghosts,
since the former is related to 2nd order operators while the latter is associated with 4th order operators.
The evolution equation of effective action with respect to the renormalization group ``time $t\equiv \ln k$'' can be expressed as
\bea\label{6dsum}
\partial_t\Gamma_k&=&3\sum_nd_n^{h^{TT}}\theta(k^6-\lambda^{h^{TT}}_n)
+3\sum_n'd_n^{\xi^{T}}\theta(k^6
-\lambda^{\xi^{T}}_n)+3\sum''_nd_n^{\sigma}\theta(k^6-\lambda^{\sigma}_n)
\nn\\&&-2\sum_nd_n^{C^T}\theta(k^2-\lambda^{C^T}_n)-2\sum'_nd_n^{C}\theta(k^2-\lambda^{C}_n)-4\sum_nd_n^{\omega^T}\theta(k^4-\lambda^{\omega^T}_n),\nn\\
&&-4\sum'_nd_n^{\omega}\theta(k^4-\lambda^{\omega}_n)+2\sum_nd_n^{f^T}\theta(k^4-\lambda^{f^T}_n)+2\sum'_nd_n^{f}\theta(k^4-\lambda^{f}_n),
\eea
where single ``prime" for $\xi$ states that the vector modes corresponding to the Killing vectors are excluded.
Single ``prime" for scalars denotes that the constant scalar modes should not be taken into account. Double ``prime" for $\sigma$ means both the constant modes and the modes corresponding to conformal Killing vectors are not taken into account.
\subsection{Evaluation of Functional Traces}

In six dimensions, $\partial_t\Gamma_k$ should take the following general form
\be
\partial_t\Gamma_k=\biggl({\cal A}_0(k)R^{-3}+{\cal A}_2(k)R^{-2}+{\cal A}_4(k)R^{-1}+{\cal A}_6+{\cal O}(R)\biggr).
 \ee
By using the spectra of harmonics on $S^6$ (See Table VI in Appendix C), we obtained
\be\label{1divs6}
\partial_t\Gamma_k=\frac{2700k^6}{R^3}-\frac{1470k^4}{R^2}-\frac{19415k^2}{54R}-\frac{601}{21}+{\cal O}(k^{-2}).
\ee
Here we emphasize that splitting a sum over eigenvalues of a 6th order operator into
sums over eigenvalues of three 2nd order operators which factor the 6th order operator will give result differing
from the original one by terms with both negative and positive powers of $k$. For instance, on $S^6$
\be
3\sum_nd_n^{h^{TT}}\theta(k^6-\lambda^{h^{TT}}_n)
=\sum_{i=1}^3\sum_nd_n^{h^{TT}}\theta(k^2-\lambda^{h^{TT}}_{n,i})
-\frac{98k^2}{3R}+{\cal O}(k^{-2}).
\ee
However, one can straightforwardly prove that ${\cal A}_0(k)$ is intact when splitting a sum over eigenvalues of a 6th order operator into three
sums over eigenvalues of three 2nd order operators which factor the 6th order operator. Meanwhile, if we denote
a $2p$th order operator by $\Delta_1$, a $2q$th order operator by $\Delta_2$ and corresponding eigenvalues
by $\lambda_1$, $\lambda_2$, we find by using the property of generalized zeta function, that the $k$ independent terms in $(p+q)\sum_n\theta(k^{2(p+q)}-\lambda_{n,1}\lambda_{n ,2})$ and $p\sum_n\theta(k^{2p}-\lambda_{n ,1})+q\sum_n\theta(k^{2q}-\lambda_{n, 2})$ are equal. With this observation, we find
that the $k$ independent term in Eq.(\ref{6dsum}) is equal to that in a simpler expression only consisting of sums over eigenvalues of 2nd order operators. The sums over eigenvalues of a 2nd order operator can be computed from Seeley-DeWitt coefficients \cite{Gilkey}. By this easier way, we obtain the ${\cal A}_6$ coefficients in Eq.(\ref{6dsum}) on symmetric Einstein spaces including $S^6$, $CP^3$, $S^2\times S^4$, $S^2\times CP^2$, $S^3\times S^3$ and $S^2\times S^2\times S^2$. These results can be found in table 3. One can also derive them by using the spectra of
Hodge-de Rahm  and Lichnerowicz operators on above backgrounds (see Table VII in Appendix D)\footnote{The spectra of Hodge-de Rahm and Lichnerowicz operators on $S^3$ can be found in appendix of \cite{Percacci:2010yk}}.

\begin{table}[ht]
\centering
\begin{tabular}{|c|c|c|c|c|c|c|}
  \hline
 Manifold &Vol/$(4\pi)^3$   & $I_1$&$I_2$&$I_3$&$E_6$&${\cal A}_6$ \\
  \hline\hline
 $S^6$   &$\frac{450}{R^3}$    &   0&0 &0&$\frac{16}{75}R^3$&   -$\frac{601}{21}$        \\
  \hline
 $CP^3$   &$\frac{576}{R^3}$   & $- \frac{29}{7200}R^3$ & $ \frac{31}{1800}R^3$ & -$ \frac{4}{75}R^3$&$\frac{2}{3}R^3$&   -$\frac{58526}{2625}$        \\
  \hline
  $S^2\times S^4$  &$\frac{324}{R^3}$     &  -$\frac{34}{6075}R^3$  &$ \frac{104}{6075}R^3$ &-$\frac{128}{2075}R^3$&$\frac{16}{27}R^3$&    -$\frac{79462}{7875}$      \\
  \hline
  $S^2\times CP^2$   & $\frac{243}{R^3}$     & -$\frac{44}{6075}R^3$  &$\frac{214}{6075}R^3$  &-$\frac{208}{2025}R^3$ &$\frac{32}{27}R^3$  &   -$\frac{11168}{2625}$  \\
  \hline
  $S^3\times S^3$   & $\frac{108\pi}{R^3}$    & -$\frac{3}{400}R^3$ & $\frac{1}{300}R^3$  &-$\frac{4}{75}R^3$ &$0$  &   -$\frac{1417\pi}{500}$  \\
   \hline
  $S^2\times S^2\times S^2$   & $\frac{216}{R^3}$   &  -$\frac{2}{225}R^3$ & $\frac{4}{75}R^3$ &-$\frac{32}{225}R^3$ &$\frac{16}{9}R^3$  &   -$\frac{732}{875}$  \\
  \hline
\end{tabular}
\caption{In this table, we present the values of conformal invariants and Euler density on $S^6$, $CP^3$, $S^2\times S^4$, $S^2\times CP^2$, $S^3\times S^3$ and $S^2\times S^2\times S^2$. The values of ${\cal A}_6$ coefficients on above backgrounds are also given. }
\label{Table3}
\end{table}

Upon integrating $\partial_t\Gamma_k$ to certain UV cutoff $\Lambda$, we can obtain the power-law and logarithmic
divergences in one-loop effective action. The logarithmic divergence can be parameterized as
\be\label{6dlog}
\Gamma_{\mbox{log}\Lambda}=\frac{\log \Lambda}{(4\pi)^3}\int d^6x\sqrt{g}\biggl(a E_6+ c_1 I_1+c_2 I_2+c_3 I_3+d\nabla_{\mu}J^{\mu}\biggr).
\ee
When evaluated on $S^6$, $CP^3$, $S^2\times S^4$, $S^2\times CP^2$, $S^3\times S^3$ and $S^2\times S^2\times S^2$, Eq.(\ref{6dlog}) should reproduce the corresponding ${\cal A}_6$ coefficients. Because $\nabla_{\mu}J^{\mu}$ vanishes on symmetric Einstein spaces, using results of Table 3, we find

\be
a=-\frac{601}{2016},\quad c_1=\frac{5633}{105}-4c_2,\quad c_3=-\frac{35543}{5040}+\frac{5c_2}{8},
\ee
That $c_i (i=1,2,3)$ can not be completely fixed is because of the following equality satisfied
on symmetric Einstein spaces.
\be
R^{\mu\nu}_{~~\sigma\rho}R^{\sigma\rho}_{~~\lambda\delta}R^{\lambda\delta}_{~~\mu\nu}=
-4R^{\mu~\nu}_{~\sigma~\rho}R^{\sigma~\rho}_{~\lambda~\delta}R^{\lambda~\delta}_{~\mu~\nu}
+\ft{1}{3}RR^{\mu\nu}_{~~\sigma\rho}R^{\sigma\rho}_{~~\mu\nu}.
\ee
Taking into account above equality, we find that in $\sum_ic_iI_i$, the coefficients of $R^{\mu~\nu}_{~\sigma~\rho}R^{\sigma~\rho}_{~\lambda~\delta}R^{\lambda~\delta}_{~\mu~\nu}$ and $RR^{\mu\nu}_{~~\sigma\rho}R^{\sigma\rho}_{~~\mu\nu}$ take definite values. They are given by $-\frac{5633}{105}$ and $\frac{20389}{18900}$ respectively. To obtain these two numerical values, we have expanded Weyl tensor in terms of Rieman tensor, Ricci tensor and Ricci scalar as well as replacing $R_{\mu\nu}$ by $\frac{R}{6}g_{\mu\nu}$.

\section{Conclusions}
Employing the Exact Renormalization Group Equation and background field method, we explore the one-loop
problem in a six-dimensional conformal gravity theory whose Lagrangian takes the same form as holographic Weyl anomaly of multiple coincident M5-branes. The backgrounds are chosen to be the symmetric Einstein spaces including $S^6$, $CP^3$, $S^2\times S^4$, $S^2\times CP^2$, $S^3\times S^3$ and $S^2\times S^2\times S^2$. By evaluating the functional sums, we obtain power-law and logarithmic divergences. The power-law divergences do not appear in \cite{Fradkin:1981iu,deBerredoPeixoto:2003pj}, because they are annihilated by dimensional regularization. Since any Einstein metric solves the equations of motion of six-dimensional conformal gravity considered in this paper, the one-loop calculation is on-shell and therefore according to \cite{Buchbinder} the logarithmic divergence should not depend on
the explicit form of gauge fixing action. Because an equality satisfied on symmetric Einstein spaces relates
different cubic curvature structures to each other, there are only three independent cubic curvature structures
$R^3$, $R^{\mu~\nu}_{~\sigma~\rho}R^{\sigma~\rho}_{~\lambda~\delta}R^{\lambda~\delta}_{~\mu~\nu}$ and $RR^{\mu\nu}_{~~\sigma\rho}R^{\sigma\rho}_{~~\mu\nu}$ on symmetric Einstein spaces. As a consequence, we cannot
fix the coefficients in front of four non-trivial conformal invariants in the one-loop logarithmic divergence. However, if the backgrounds are chosen to be the more general Einstein spaces, we will not achieve the maximal factorization in the gauge fixed quadratic action, and the calculation becomes infeasible. From the explicit values of logarithmic divergence on above backgrounds,
we extract the coefficient in front of Euler density and two linear equations constraining the coefficients
in front of three type-B conformal invariants. It should be interesting and challenging to conceive a new way by which the three coefficients in front of type-B conformal invariants can be entirely fixed. Direct application of generalized Schwinger-DeWitt method to six dimensions would be extremely tedious.

\section*{Acknowledgement}
Very useful discussions with R. Percacci, C. N. Pope and E. Sezgin are gratefully acknowledged. We also thank communications with S. Fulling. This work was supported in part by DOE grant DE-FG03-
95ER40917i.

\appendix

\section{Spectrum of Laplacian Operator on $S^4$}
\begin{table}[ht]
\centering
\begin{tabular}{|c|c|c|c|c|}
  \hline
  Spin &Operator & Eigenvalue  & Degeneracy& Lowest value of $n$   \\
  \hline\hline
 0   &-$\Box$&$\frac{R}{12}(n^2+3n)$  & $\frac{1}{6}(n+1)(n+2)(2n+3)$     &    0            \\
  \hline
  $1$   &-$\Box$     & $\frac{R}{12}(n^2+3n-1)$    &$\frac{3}{6}n(n+3)(2n+3)$ &  1           \\
  \hline
   2       &   -$\Box$       & $\frac{R}{12}(n^2+3n-2)$  & $\frac{5}{6}(n-1)(n+4)(2n+3)$ & 2    \\
  \hline
\end{tabular}
\caption{In this table, we present the spectra of scalar, vector and tensor harmonics on $S^4$. }
\label{Table4}
\end{table}

\section{Spectra of Hodge-de Rahm and Lichnerowicz operators on $CP^2$}
\begin{table}[ht]
\centering
\begin{tabular}{|c|c|c|c|c|}
  \hline
  Spin &Operator & Eigenvalue  & Degeneracy& Lowest value of $n$   \\
  \hline\hline
 0   &-$\Box$&$\frac{R}{6}(n^2+2n)$  & $(n+1)^3$     &    0            \\
  \hline
  $1$   &-$\Box+\frac{R}{4}$     & $\frac{R}{6}(n^2+2n)$    &$(n+1)^3$ &  1           \\
  \hline
  $1$   &-$\Box+\frac{R}{4}$     & $\frac{R}{6}(n^2+5n+6)$    &$(n+1)(n+4)(2n+5)$ &  0           \\

  \hline
   2       &   $\Delta_L$       & $\frac{R}{6}(n^2+2n)$  & $(n+1)^3$ & 2    \\
  \hline
   2       &   $\Delta_L$       & $\frac{R}{6}(n^2+5n+6)$  & $(n+1)(n+4)(2n+5)$ & 1   \\
    \hline
   2       &   $\Delta_L$       & $\frac{R}{6}(n^2+8n+18)$  & $(n+1)(n+7)(2n+8)$ & 0   \\
   \hline
\end{tabular}
\caption{In this table, the spectra of Hodge-de Rahm and Lichnerowicz operators related to scalar, vector and tensor on $CP^2$ are given. An explicit construction of these harmonics can be found in \cite{Pope:1980ub,Pope:1982ad} and a group theoretical way of
obtaining the spectra has been discussed in \cite{Vassilevich:1993yt} }
\label{Table5}
\end{table}
\section{Spectrum of Laplacian Operator on $S^6$}
\begin{table}[ht]
\centering
\begin{tabular}{|c|c|c|c|c|}
  \hline
  Spin &Operator & Eigenvalue  & Degeneracy& Lowest value of $n$   \\
  \hline\hline
 0   &-$\Box$&$\frac{R}{30}(n^2+5n)$  & $\frac{1}{120}(n+1)(n+2)(n+3)(n+4)(2n+5)$     &    0            \\
  \hline
  1   &-$\Box$     & $\frac{R}{30}(n^2+5n-1)$    &$\frac{5}{120}n(n+2)(n+3)(n+5)(2n+5)$ &  1           \\
  \hline
   2       &   -$\Box$       & $\frac{R}{30}(n^2+5n-2)$  & $\frac{14}{120}(n-1)(n+2)(n+3)(n+6)(2n+5)$ & 2    \\
  \hline
\end{tabular}
\caption{In this table, we exhibit the spectra of scalar, vector and tensor harmonics on $S^6$.}
\label{Table6}
\end{table}

\section{Spectra of Hodge-de Rahm and Lichnerowicz operators on $CP^3$}
Since $CP^3\simeq SU(4)/(SU(3)\times U(1))$ is a coset space, one can use the method presented in \cite{Pilch:1984xx} to calculate the spectrum.
\begin{table}[ht]
\centering
\begin{tabular}{|c|c|c|c|c|}
  \hline
  Spin &Operator & Eigenvalue  & Degeneracy& Lowest value of $n$   \\
  \hline\hline
 0   &-$\Box$&$\frac{R}{12}(n^2+3n)$  & $\ft{1}{12}(1+n)^2(2+n)^2(3+2n)$     &    0            \\
  \hline
  $1$   &-$\Box+\frac{R}{6}$     & $\frac{R}{12}(n^2+3n)$    &$\ft{1}{12}(1+n)^2(2+n)^2(3+2n)$ &  1           \\
  \hline
  $1$   &-$\Box+\frac{R}{6}$     & $\frac{R}{12}(n^2+6n+8)$    &$\ft{8}{12}(1 + n) (3 + n)^3 (5 + n)$ &  0           \\

  \hline
   2       &   $\Delta_L$       & $\frac{R}{12}(n^2+3n)$  & $\ft{1}{12}(1+n)^2(2+n)^2(3+2n)$ & 2    \\
  \hline
   2       &   $\Delta_L$       & $\frac{R}{12}(n^2+6n+8)$  & $\ft{8}{12} (1 + n) (3 + n)^3 (5 + n)$ & 1   \\
    \hline
   2       &   $\Delta_L$       & $\frac{R}{12}(22 + 9 n + n^2)$  & $\ft{6}{12}(1 + n) (4 + n) (5 + n) (8 + n) (9 + 2 n)$ & 0   \\
   \hline
    2       &   $\Delta_L$       & $\frac{R}{12}(n^2+5n+6)$  & $\ft{3}{12} (1 + n)^2 (4 + n)^2 (5 + 2 n)$ & 0   \\
    \hline
\end{tabular}
\caption{In this table, we present the spectra of Hodge-de Rahm and Lichnerowicz operators on $CP^3$. }
\label{Table7}
\end{table}

\end{document}